\documentclass[12pt,notoc,article]{JHEP3}

\usepackage{amsmath,amssymb,euscript,array,mathrsfs}

\usepackage{epsfig}

\def\N{{\cal N}}

\def\Tr{{\rm Tr}}

\def\det{{\rm det}}

\def\S{S}

\newcommand{\Dslash}{D\mkern-11.5mu/\,} 
\newcommand{\delslash}{{\partial\mkern-9mu/}}

\def\Dbarslash{\,\,{\raise.15ex\hbox{/}\mkern-12mu {\bar D}}}
\def\Dslash{\,\,{\raise.15ex\hbox{/}\mkern-12mu D}}
\def\delslash{\,\,{\raise.15ex\hbox{/}\mkern-9mu \partial}}
\def\delbarslash{\,\,{\raise.15ex\hbox{/}\mkern-9mu {\bar\partial}}}

\newcommand{\MAT}[1]{\begin{pmatrix} #1\end{pmatrix}}
\newcommand{\EQ}[1]{\begin{equation} #1 \end{equation}}
\newcommand{\AL}[1]{\begin{subequations}\begin{align} #1
\end{align}\end{subequations}}
\newcommand{\SP}[1]{\begin{equation}\begin{split} #1
\end{split}\end{equation}}

\title{$\mathbf{{\cal N}=4}$ SYM on $\mathbf{S^3}$ with Near Critical Chemical Potentials}
\author{Timothy J. Hollowood, S. Prem Kumar,
  Asad Naqvi and Philip Wild \\

{\it Department of Physics,\\ Swansea University,\\
Swansea, SA2 8PP, UK.\\\\

} \\\\
E-mail: {\tt t.hollowood,s.p.kumar,a.naqvi,pypw@swan.ac.uk}}
\abstract{ We study the ${\cal N}=4$ theory at weak coupling, 
on a three sphere in the grand canonical ensemble 
with R symmetry chemical potentials. We focus attention on near
critical values for the chemical potentials, above which the
classical theory has no ground state. By computing a one loop
effective potential for the light degrees of freedom in this regime,
we show the existence of flat directions of complex dimension $N$, $2N$
and $3N$ for one, two and three critical chemical potentials
respectively; these correspond to one half, one quarter and one-eighth
BPS states becoming light respectively at the critical values. 
At small finite temperature we show that the chemical
potentials can be continued beyond their classical limiting values to
yield a deconfined metastable phase with lifetime diverging in the
large $N$ limit. Our low temperaure analysis complements the high
temperature metastability found by Yamada and Yaffe. The resulting 
phase diagram at weak coupling bears a striking resemblance to the
strong coupling phase diagram for charged AdS black holes. Our
analysis also reveals subtle qualitative differences between the two regimes.}

\begin{document}

\section{Introduction}

Among the many deep insights to emerge from the AdS/CFT
correspondence \cite{Maldacena:1997re, Gubser:1998bc,Witten:1998qj}, is the
remarkable connection between large $N$ Yang-Mills thermodynamics at strong
coupling and black holes in $AdS$ spacetimes. 
A dramatic consequence of
this equivalence is that the first order
Hawking-Page transition in string theory on asymptotically
$AdS_5\times S^5$ geometries  
corresponds to a deconfinement phase transition in the  
${\cal N}=4$ supersymmetric, large $N$ Yang-Mills
theory on the conformal boundary of the spacetime, namely on $S^3\times
S^1$ \cite{Witten:1998zw}. Although the above picture most naturally applies 
at strong 't Hooft coupling,                                              
it is now well appreciated that such a first order deconfinement
transition  occurs even in the large $N$ {\em free} Yang-Mills theory
on $S^3\times 
S^1$, with
thermal boundary 
conditions \cite{Sundborg:1999ue,Aharony:2003sx,Aharony:2005bq}. It is
believed that the behaviour 
might possibly extend to the weakly interacting theory. 

This raises the exciting possibility that thermodynamics of the weakly
coupled ${\cal N}=4$ theory on $S^3$ could be qualitatively similar to
the strongly interacting case and may provide a window into the
physics of black holes in string theory. The outstanding issue is to
understand how  the regimes of weak and strong 't Hooft coupling are
mapped into one another as the coupling is changed. 
Of course, it is still 
plausible that, away from the extreme limits of 
  infinite and zero 't Hooft couplings, the phase structure at
  non-zero weak
  coupling could be  qualitatively
  different from that at finite strong coupling. 
This latter possibility, if true, would imply
  non-analyticities in the theory as a function of the 't Hooft
  coupling, and qualitatively different thermodynamics of string
  theory on strongly curved backgrounds 
\cite{Gao:1998ww,Hartnoll:2005ju} compared to semiclassical gravity. 

For these reasons, mapping out the thermodynamic phase structure of
${\cal N}=4$ theory has received much attention from various
viewpoints in recent years
\cite{Liu:2004vy,Yamada:2006rx,Harmark:2006di,Hollowood:2006xb,
Harmark:2006ta,Harmark:2006ie,Harmark:2007px,AlvarezGaume:2005fv,
Basu:2005pj,AlvarezGaume:2006jg,Azuma:2007fj,Dutta:2007ws}.   

In this paper we study the weakly coupled ${\cal N}=4$ theory with
$SU(N)$ gauge group,
on an $S^3$ of radius $R$, in the grand
  canonical ensemble with chemical potentials for
global R-symmetry charges. The ${\cal N}=4$ theory
has three global $U(1)$ symmetries generated by the Cartan elements of
the $SU(4)$ R-symmetry. 
To pass to the
grand canonical ensemble, we introduce chemical potentials
$(\mu_1,\mu_2,\mu_3)$ for the 
three global R-symmetry charges $(J_1,J_2,J_3)$ in the theory.

We will explore a particularly interesting corner of the
phase diagram which turns out to be analytically tractable in the
presence of chemical potentials. When the ${\cal N}=4$ theory is
formulated on the three sphere of radius $R$, the six scalar fields 
obtain a mass $1/R$ due to conformal coupling to the curvature of
the sphere. In addition, the introduction of a chemical potential for
a global charge induces an effective 
negative mass squared for scalars carrying that charge. There is
thus a critical regime of values for the chemical potentials
$\mu_p=R^{-1} +{\cal O} (\lambda)$, 
for which the scalars become light degrees of freedom. (Here $\lambda\ll1$ 
is the 't Hooft coupling.) We will refer to this region as the
``near critical'' regime.

The classical theory is only stable
when $\mu_p\leq R^{-1}$, and this statement is true also at
one loop order. When the chemical potential exceeds this
value, the classical Hamiltonian becomes unbounded from below. Note
that in a simpler field theory, such as a massive complex
scalar field  theory with a quartic interaction, introducing a
chemical potential exceeding the mass leads to Bose-Einstein
condensation. This does not appear to be the case, at least
perturbatively, in the ${\cal N}=4$ theory where the quantum corrections
at weak coupling are always systematically smaller than the tree level
mass term. The only situation 
where there is a possibility for tree level and one loop radiative
corrections to compete is when $\mu_p\simeq R^{-1} +{\cal O}
(\lambda)$, so that the classical potential already contains a term of
order $\lambda$.

In addition to requiring near-criticality
of chemical potentials, for the most part we will focus on 
low temperatures $TR\ll1$. Our analysis  
complements the work of Yamada and Yaffe  
\cite{Yamada:2006rx} who investigated the weakly interacting theory
\footnote {The authors of \cite{Yamada:2006rx} also studied the {\em free
  theory} in the grand canonical ensemble, obtaining a line of 
   first order Hagedorn/deconfinement transitions for generic values of
the chemical potential. We do not address the issue as to whether
or not 
the first order line gets modified at non-zero weak coupling.} 
in the temperature range ${1\ll TR \lesssim1/\sqrt{\lambda}}$. One of their 
interesting results was to demonstrate the existence of a metastable
deconfined plasma phase, at high temperatures in the
range $R^{-1}\leq \mu_p<\sqrt{\lambda T^2+{R}^{-2}}$.

In the near critical regime where two or more scalars are light,
there exist classically (almost) flat directions. They are
parametrized by mutually commuting, constant background values for the
light scalars. At a small finite temperature, the shallow directions include 
mutually commuting constant configurations for the thermal
Wilson-Polyakov line. 
Moving along
these directions generically Higgses the theory: $SU(N)\rightarrow
U(1)^{N-1}$. For sufficiently large values of the diagonal modes of
the scalars, the off-diagonal modes of all the Kaluza-Klein harmonics
acquire large masses. These can be integrated
out to generate a Wilsonian effective potential at one loop for the 
light, diagonal modes in the regime of near critical chemical potentials.

Our first result in the near critical regime at $T=0$ is that
the one loop radiative correction to the scalar potential vanishes,
following a non-trivial Bose-Fermi cancellation in regularized Casimir
sums in the  presence of expectation values for the light scalar
modes. This means that the nearly flat directions (at $T=0$) are not
lifted by quantum corrections in the vicinity of critical chemical potentials.

In particular, when the chemical potentials are {\em at} 
their critical values, and at $T=0$, due to vanishing quantum
corrections there is a Coulomb branch moduli space of complex
dimension $N$, $2N$ or $3N$, depending on whether we have one, two
or three critical chemical potentials switched on. For each case we
also have a different number of zero energy modes: With
$\mu_1=R^{-1},\mu_2=\mu_3=0$ a single {\em holomorphic} 
adjoint scalar mode with zero energy appears, while for  
$\mu_1=\mu_2=R^{-1},\mu_3=0$, there are $2$ holomorphic scalar
zero modes. The situation with 
three critical chemical potentials reveals 
two adjoint fermion zero energy modes along with $3$
holomorphic scalar zero modes.

The appearance of the moduli spaces at critical chemical potential and
the associated zero energy modes can be understood more generally as follows. 
The generator of time translations of the 
${\cal N}=4$ theory with chemical potentials on $S^3$, may be 
expressed as
\EQ
{\Delta(\mu_p)=\Delta -\sum_{p=1}^3\mu_p\,J_p,
}
where $\Delta$ is identified with
the dilatation operator of the theory formulated on ${\mathbb
  R}^4$. 
With one critical chemical potential $\mu_1=R^{-1}$ (and $\mu_2=\mu_3=0$),
this operator vanishes on all states with $R\Delta=J_1$, which are the
infinite set of ${1\over 2}$ BPS operators in the theory. At the
critical values for $\mu_p$, the
superconformal algebra ensures positive definiteness of the above Hamiltonian
and the ${1\over 2}$ BPS operators thus constitute the infinitely
degenerate set of ground states of the theory. This can be interpreted
as the origin of the flat directions at critical chemical
potential. For two and three critical chemical potentials, the ground
states are parametrized by ${1\over 4}^{\rm th}$ and ${1\over 8}^{\rm
  th}$ BPS operators respectively. The dimensions of the moduli spaces
we find and associated zero modes are consistent with expectations based
on our knowledge of the generators 
of ${1\over 2}$ BPS, ${1\over 4}^{\rm th}$ BPS and
${1\over 8}^{\rm th}$ BPS states in the ${\cal N}=4$ theory
\cite{Kinney:2005ej, Dolan:2007rq}.

At the critical values for the chemical potentials, upon switching on a 
small non-zero temperature $TR\ll 1$ the theory acquires another set
of zero modes in addition to the light scalars. These new zero
modes are the diagonal elements of the Polyakov loop matrix. We find a
joint effective potential for all the light modes and deduce that
eigenvalues of the Polyakov loop matrix experience purely a mutual
attractive force causing them to all collapse on to a point. This
corresponds to a deconfined phase wherein the trace of the Polyakov loop
has non-zero expecatation value. 
Furthermore, the theory develops a
mass gap due to exponentially small thermal masses at low temperature. This
means that thermal effective potential for the scalars has a positive
curvature near the origin at critical chemical potential. Away from
the origin, at large
field amplitudes it asymptotes to a constant $(3/16 R)$. The
small positve curvature near the origin allows us to raise the
chemical potentials beyond their classical limiting value $1/R$, and still
obtain a locally stable configuration at the origin. Raising the
chemical potential(s) above $1/R$, however, causes the scalar potential
to have a runaway behaviour (unbounded from below) at large field amplitudes.
This results in
a metastable state with an exponentially diverging lifetime in the
$N\to \infty$ limit. We expect this low temperature metastable phase
to be a smooth continuation of the high temperature metastable plasma 
discovered in \cite{Yamada:2006rx}. 

Technically, there is an important difference
between the high and low temperature regimes. At high temperatures,
$TR \gg 1$, the effective potential is obtained basically by a flat space
computation on ${\mathbb R}^3$. The low temperature effective potential on the
other hand, in the presence of non zero R-charge densities
$(\mu_p\neq0)$ depends on the details of the compact space
on which the theory is formulated. Our analyisis complements the work
of \cite{Yamada:2006rx}, filling in the low temperature regime of the
phase diagram of the theory with chemical potentials.
The final weak coupling phase diagram is shown in Figure 3. 

Perhaps surprisingly, a quick comparison of Figure 3 with the strong
coupling phase diagram in Figure 4 reveals striking similarities. At
strong coupling, the phase diagram is dictated by the thermodynamics
and stability properties of R-charged black holes in AdS space. More
detailed aspects of these are discussed in Section 4. Here we
further remark that it has also recently been found \cite{Yamada:2008em} 
that the region in the
strong coupling phase diagram, below the black hole instability line
and above the critical value of the chemical potential, exhibits a
metastability . This metastability corresponds to a singe (probe)
D3-brane splitting from a cluster of large $N$ rotating branes whose
near horizon geometry is the charged AdS black hole
background.\footnote{As emphasized in \cite{Yamada:2008em} this  
  phenomenon is distinct from the source of local thermodynamic 
instability found in \cite{Gubser:1998jb,Cvetic:1999ne}. 
} This
is exactly the physics expected from the weak coupling analysis where
the lifetime of the metastable state is determined by the probability
for one scalar eigenvalue to tunnel out or be thermally activated into
the unstable region. 

Our weak coupling analysis, however, also reveals certain important
differences with the strongly coupled regime. We find that, in the
$\mu-T$ plane, the metastable region shrinks to zero size at zero
temperature -- the instability line meets the first order 
deconfinement line at $\mu=1/R$ and $T=0$ for any choice of chemical
potentials. At strong coupling, the black hole instability line and
the first order Hawking-Page lines meet at $\mu=1/R$ and non-zero
temperature. Only for the case with equal chemical potentials do the
two lines meet at $T=0$ and $\mu=1/R$ at strong coupling.

The organization of this paper is as follows. In Section 2, we review
how R-symmetry chemical potentials are introduced in the ${\cal N} =4$
theory. In section 3, we show how to compute the one loop effective
potential for the light degrees of freedom near critical values
for chemical potentials. We perform the calculations with one, two and
three critical chemical potentials at zero temperature. In Sections 3.2 and 3.3
we establish the existence of flat directions at zero temperature and
their interpretation in terms of BPS states becoming light. Sections
3.4 - 3.5 are 
devoted to establishing the existence of the metastable plasma phase at
low temperatures. In Section 4, we describe the similarities and
differences between the phase diagrams at weak and strong
coupling. Conclusions and future directions are summarized in Section
5, and finally, an Appendix is devoted to the spherical harmonic
decomposition of the theory on $S^3$.

\section{R Symmetry Chemical Potentials}

The $\N=4$ theory has an $SU(4)_R$ global symmetry. There are thus 
three chemical potentials which can be introduced associated to the
maximal abelian subgroup $U(1)^3\subset SU(4)_R$. The six adjoint
scalars $\{\phi_i\}$ $(i=1,2,\ldots,6)$ transform as the antisymmetric
${\bf 6}$ of $SU(4)_R$, while the fermions are in
the fundamental representation, the ${\bf 4}$ of $SU(4)_R$. (We follow 
the conventions of \cite{Yamada:2006rx} below).

We choose the three $U(1)$ generators of the Cartan subalgebra to
be 
\SP{&R_1^{\bf 4}=\tfrac{1}{2}\,{\rm diag}(1,1,-1,-1),\\
&R_2^{\bf 4} =\tfrac{1}{2}\,{\rm diag}(1,-1,1,-1),\\
&R_3^{\bf 4} =\tfrac{1}{2}\,{\rm diag}(1,-1,-1,1),}
in the fundamental representation. Packaging the six real scalars into
three complex combinations,
\EQ{\Phi_1=\frac{1}{\sqrt
    2}(\phi_1+i\phi_2),\quad\Phi_2=\frac{1}{\sqrt2}(\phi_3+i\phi_4),
\quad\Phi_3= 
{1\over\sqrt 2}(\phi_5+i\phi_6)}
we can define the six-vector
\EQ{\vec{\Phi}=(\Phi_1,\Phi_1^*,\Phi_2,\Phi_2^*,\Phi_3,\Phi_3^*).
\label{sixv}}
The three $U(1)$ generators acting in this representation are then  
\SP{&R_1^{\bf 6}=\tfrac{1}{2}\,{\rm diag}(1,-1,0,0,0,0),\\
&R_2^{\bf 6} =\tfrac{1}{2}\,{\rm diag}(0,0,1,-1,0,0),\\
&R_3^{\bf 6} =\tfrac{1}{2}\,{\rm diag}(0,0,0,0,1,-1),}
which clearly assigns opposite charges to the fields $\Phi_i$ and
their complex conjugates. 

The grand canonical partition
function is defined to be
\EQ{
{\cal Z}(T,\mu_p)=\text{Tr} \,e^{-\beta \left(\Delta-\sum_p
\mu_p J_p\right)}\
,
\label{partfn}}
where the $J_p$ are the three associated conserved R charges. The
charge densities involve both fermionic and bosonic
contributions. 
With the above choice of the $U(1)$ generators, the chemical potential
assignments are determined via the following expressions
\EQ
{\sum_{i=1}^3 \mu_i \, R_i^{\bf 6}=(\mu_1,-\mu_1, \mu_2,-\mu_2,\mu_3,-\mu_3)}
 for the six scalars in Eq.\eqref{sixv}. Similarly, the fermion
 chemical potentials are determined as 
\EQ
{\sum_{i=1}^3\mu_i \, R_i^{\bf 4}=\,{\rm diag}(\bar\mu_1,\bar\mu_2,\bar\mu_3,\bar\mu_4)}
where
\SP{
\bar\mu_1&=\tfrac12(\mu_1+\mu_2+\mu_3)\\
\bar\mu_2&=\tfrac12(\mu_1-\mu_2-\mu_3)\\
\bar\mu_3&=\tfrac12(-\mu_1-\mu_2+\mu_3)\\
\bar\mu_4&=\tfrac12(-\mu_1+\mu_2-\mu_3)\, .
\label{ferchemi}
}

The grand canonical partition
function Eq.\eqref{partfn} can also be realized as a Euclidean
functional integral for 
the theory on $\S^3\times\S^1$. In the functional integral or
Lagrangean formulation, the chemical potential for each global 
charge can be thought of as introducing by hand, 
a constant (imaginary) background for the time component of a gauge
field associated to the respective global 
$U(1)$ symmetry. In the presence of the R charge chemical potentials the
Lagrangean for the ${\cal N}=4$ theory on $S^3\times S^1$ becomes
\SP
{&{\cal L}={1\over g^2}\Tr\left(
\tfrac{1}{2}F_{\mu\nu}F^{\mu\nu}+ \tfrac{1}{2}\sum_{p=1}^3(D_\nu
\phi_{2p-1} -i\mu_p\delta_{\nu, 0}\phi_{2p})^2+\tfrac{1}{2}\sum_{p=1}^3(D_\nu
\phi_{2p} +i\mu_p\delta_{\nu, 0}\phi_{2p-1})^2
+\right.\\
&\left.{1\over 2R^2}\phi_a^2-{1\over 2}
  [\phi_a,\phi_b]^2+
i\bar \psi_A
  \left(\Dslash-\bar\mu_A\gamma_0\gamma_5\right)\psi_A - \bar\psi_A\left[
\alpha_{AB}^p\phi_{2p-1}+i\beta^p_{AB}\gamma_5\phi_{2p},
\,\psi_B\right]\right)
\label{lag}}
where all derivatives are gauge covariant derivatives on $S^3$. The
$\psi^A$ are four component Majorana fermions
\EQ
{\psi^A = \left( \lambda_\alpha^A,\,\bar\lambda^{A\dot\alpha}\right)^T
}

The indices
$a,b=1,\ldots 6$, and $A,B=1,\ldots 4$, while $p=1,2,3$. The $4\times
4$ matrices, $\alpha^p$ and $\beta^p$ are the Clebsch-Gordan
coefficients, satisfying
\EQ{\{\alpha^p,\alpha^q\}=-2
  \delta^{pq},\quad\{\beta^p,\beta^q\}=-2\delta^{pq},
  \quad[\alpha^p,\beta^q]=0.}  
Explicit representations for the $\alpha$ and $\beta$ matrices are 
given in terms of Pauli matrices as,
\AL
{&\alpha^1=\left( \begin{array}{cc}
i\sigma_2 & 0  \\
0 & i\sigma_2 \end{array} \right),\quad
\alpha^2=\left( \begin{array}{cc}
0 & -\sigma_3  \\
\sigma_3 & 0 \end{array}\right),\quad\alpha^3= \left( \begin{array}{cc}
0 & \sigma_1  \\
-\sigma_1 & 0 \end{array}\right)\\
&\beta^1=\left( \begin{array}{cc}
-i\sigma_2 & 0  \\
0 & i\sigma_2 \end{array} \right),\quad
\beta^2=\left( \begin{array}{cc}
0 & \sigma_0  \\
-\sigma_0 & 0 \end{array}\right),\quad
\alpha^3= \left( \begin{array}{cc}
0 & i\sigma_2  \\
i\sigma_2 & 0 \end{array}\right)}
\\
\underline{\it Classical (in)stability:}

It is immediately clear from the scalar kinetic terms in 
\eqref{lag} that in the presence of
chemical potentials 
the Euclidean action is not real.
Notice also that with the chemical potential, the conformal scalars have an
effective mass given by 
\EQ
{m^2_{p}=R^{-2}-\mu_p^2\qquad p=1,2,3.}

Hence
only when $m^2_{p}\geq 0$, does the classical theory have a stable
vacuum. For 
$\mu_p>R^{-1}$ there is a classical instability in the theory along
directions in field space for which $\phi_{2p-1}$ and $\phi_{2p}$
commute. It follows that in flat space, {\it i.e.} on ${\mathbb R}^4$,
where all the fields are exactly massless, it
is not possible to introduce a chemical potential in ${\cal N}=4$
SYM since the resulting theory has no ground state (at least
classically) and the grand canonical ensemble is ill-defined. In
finite volume and in particular on $S^3$, the conformal coupling of
the scalars to the background curvature allows for a mass  which
leads to a stable vacuum for a range of values of the chemical potential.

It is worth noting that this kind of instability 
in a massive {\it interacting} scalar field theory, driven by a chemical
potential exceeding the mass, generally leads to Bose-Einstein
condensation, {\it i.e.} a VEV for the scalar fields and a spontaneous
breaking of the $U(1)$ symmetry.  
For example, this occurs in the $\phi^4$ theory, where the
interactions stabilize the ground state 
at some finite VEV. In the $\N=4$ theory, classically at least there
is nothing to stabilize the theory along the mutually commuting
directions of configuration space when 
one or more of the $\mu_p$ exceed $R^{-1}$. Physically,
 when $\mu_p>R^{-1}$ the system can always lower the energy by
populating the vacuum with any number of charged quanta which it can
borrow from the bath. 

In the 
vicinity of the critical chemical potentials $\mu_p\sim
R^{-1}$, the scalar fields of the theory are the light, almost massless
degrees of freedom. The appearance of these new light modes makes the
approach to the critical chemical potential an interesting regime to study.
In this paper we consider these near-critical regions in more
detail. There are three cases which will each be considered
separately: 
\SP{
&{\rm (i)}\quad\mu_1\simeq R^{-1},\quad\mu_2=\mu_3=0;\\\\
&{\rm (ii)}\quad\mu_1\simeq\mu_2\simeq R^{-1},\quad\mu_3=0;\\\\ 
&{\rm (iii)}\quad\mu_1\simeq\mu_2\simeq\mu_3\simeq R^{-1}. 
\label{cases}}
%
We will consider the approach to the critical chemical,
with vanishing as well as
small non-zero temperatures.

\section{One Loop Effective Potential -  Generalities}

In this  section of the paper, we
set up the 
calculation of a one-loop effective potential for the $\N=4$ theory on
$\S^3$ with a radius $R$, at both zero and non-zero (but small)
temperatures in each of the near-critical regions \eqref{cases} above.

Adopting a Wilsonian approach, we will compute the effective
potential for the lightest degrees of freedom in theory, by
integrating out all the other heavy modes in the background of the
light modes. The natural mass scales for most of the heavy degrees of
freedom are the inverse
radius of the $S^3$, namely $R^{-1}$
and the temperature $T=\beta^{-1}$. All the fields can be expanded in
terms of spherical harmonics on $\S^3$ and
Matsubara modes on the thermal circle. The light modes must
necessarily be constant modes on $\S^3\times \S^1$. The only fields
which have such constant modes are the scalars $\phi_a$ and $A_0$, the gauge
field component around $\S^1$.\footnote{The other gauge field
  components $A_i$ are vector-valued on $\S^3$ and so do not have a
  constant mode.} The scalars 
generally have a mass of order $R^{-1}$ since they are conformally
coupled to the background curvature. 
However, as we have seen above, in the presence
of a near-critical chemical potential, $\mu_p\simeq
R^{-1}$, the effective masses of $\phi_{2p-1}$ and $\phi_{2p}$
are small and so we should include their constant
modes in the effective potential. 

Thus there are two sets of light modes in the near critical
theory. First, we have for each near-critical $\mu_p$, light
adjoint scalar modes which are the homogeneous parts of the respective
fields on $S^3$:
\EQ
{
\varphi_{2p-1}= {T\over 2\pi^2 R^3}\int_{S^3\times
    S^1}\phi_{2p-1}\,,\qquad
\varphi_{2p}= {T\over 2\pi^2 R^3}\int_{S^3\times
    S^1}\phi_{2p}\,;\qquad\mu_{p}\simeq {R}^{-1}.\label{varphi}}
In addition to these, at finite temperature, we must also account for
the spatial zero mode of the holonomy of the time component of the
gauge field around the thermal circle:
\EQ
{\alpha = {T\over 2\pi^2 R^3}\int _{S^3\times S^1} A_0.}

The Wilsonian effective potential has a tree-level contribution, as well as
loop corrections,
\EQ{ V_\text{eff}=\lambda^{-1}V_0+V_1+{\cal
    O}(\lambda).\label{pertpot}}
Here $\lambda =g^2 N$ is the 't Hooft coupling. Although we only
restrict to one loop computations, we will consistently express all
quantities as fucntions of the 't Hooft coupling and $N$, indpendently,
since these are the parameters relevant for evenetual 
comparison with the large $N$ gravity dual. 
The tree-level term is
\EQ{
V_0=
\frac{N\pi^2R^3}
{2\lambda}\text{Tr}\,\Big(-\sum_{a}[A_0,\phi_a]^2
-\sum_{a<b}[\phi_a,\phi_b]^2
+\sum_p(R^{-2}-\mu_p^2)(\phi_{2p-1}^2+\phi_{2p}^2)\Big)\ .
}
When the chemical potentials are small so that $\mu_p<R^{-1}$,  the 
tree-level potential forces the associated scalar
VEVs to vanish. On the other hand, if $\mu_p>R^{-1}$,  the theory
becomes unstable. In the region of one or more
near-critical chemical potentials,
$\mu_p\simeq R^{-1}$, the classical potential has almost flat
directions. These flat directions correspond to
mutually commuting VEVs $\varphi_{2p-1}$ and
$\varphi_{2p}$, in addition to a mutually commuting $\alpha$.

\subsection*{One loop correction near-critical chemical potentials}

We will now outline the computation of 
the one-loop effective potential in the theory
approaching the critical chemical potential in
 each of the three regions \eqref{cases}. 
The near-critical region for the chemical potentials will be defined
specifically as  
\EQ
{\mu_p=R^{-1}+{\cal O}(\lambda).\label{nearcrit1}
}
This parametrically small approach towards criticality is chosen
so that the tree-level potential for the commuting modes is of 
the same order as the first loop correction. It is in this situation
that there is the possibility of competition between these two effects with
potentially interesting physics. If $|\mu_p-R^{-1}|$ is parametrically
larger, then depending on the sign, the theory will simply either have
a stable vacuum at the origin, or have perturbatively unstable runaway
directions.

For each critical chemical potential $\mu_p$, classically the theory will have
almost flat directions parametrized by (a holomorphic combination of) 
the diagonal 
elements of mutually commuting matrices $\varphi_{2p-1}$ and
$\varphi_{2p}$, $(p=1,2,3)$. In addition, there is another set of
moduli parametrized by the diagonal elements of $\alpha$, which 
also commutes with $\varphi_{2p}$ and $\varphi_{2p-1}$ at a minimum of
the classical potential. Therefore at a generic point along this flat
potential we have,
\EQ
{
\alpha ={\rm diag}(\alpha_i)\qquad\varphi_{2p-1}= {\rm
  diag}\left(\varphi_{2p-1 \,i}\right),\qquad\varphi_{2p}= 
{\rm diag}\left(\varphi_{2p \,i}\right),\qquad{i=1,2,\ldots,N}.
}
Recall from \eqref{varphi} that the fields 
$(\alpha,\varphi_a)$ are the background values for the spatially homogeneous
parts of the full quantum fields $(A_0,\phi_a)$.

\subsubsection*{Near-critical chemical potential for only one of the three U(1)s}

We begin our analysis with the simplest situation where only one
near critical chemical potential $\mu_1$ is turned on: 
\EQ
{
\mu_1\simeq R^{-1}, \mu_2=\mu_3=0.
}
In this situation, the diagonal elements of the fields 
$\varphi_1, \varphi_2$ and $\alpha$ are the lightest modes in the theory.

At generic points of the classically flat directions parametrized by
the diagonal elements of $\varphi_1$ and $\varphi_2$,
we will integrate out all the
inhomogeneous modes (the Kaluza-Klein harmonics on $S^3$) as well as
the off-diagonal homogeneous fluctuations. In order to ensure the
validity of the semiclassical or one loop approximation, it will be
necessary that the off-diagonal fluctuations have relatively large
masses. From the Lagrangian \eqref{lag}, we can estimate the masses of 
the light off-diagonal excitations of
the zero momentum scalar modes to be
\EQ
{m^2_{ij}\sim{\left(\varphi_{1, ij}^2+\varphi_{2, ij}^2\right)
+R^{-2}-\mu_1^2}}
where $\varphi_{a,ij}\equiv\varphi_{ai}-\varphi_{aj}$. Validity of 
perturbation theory in the Wilsonian sense, about diagonal scalar
backgrounds then requires that the off diagonal modes be much heavier
than the light diagonal degrees of freedom:
\EQ
{R^2\sum_p\left(\varphi_{1, ij}^2+\varphi_{2, ij}^2\right)
 \gg |1-(\mu_1 R)^2|\simeq \;{\cal O}(\lambda). 
}
This condition is easy  to ensure in the near critical
region \eqref{nearcrit1} by choosing appropriately large VEVs for all
the diagonal entries, such that the differences between them also
remain parametrically large.


The masses of all heavy excitations will depend only on the
sum
\EQ
{\varphi^2_{ij}\equiv\left(\varphi_{1,ij}^2+\varphi_{2,ij}^2\right),
\label{lal}}
 as seen
above.\footnote{In the presence of diagonal
background VEVs for the lightest scalars, {\it all} the degrees of freedom
which are integrated out are necessarily the 
off-diagonal fluctuations of every KK
harmonic on $S^3$.} For this reason, at leading order in the coupling, 
the one-loop contribution to the effective potential for light modes
only depends on this combination.
In this case, we can use
the resulting symmetry to only turn on a
VEV for one of the scalar fields, which we can take to
be $\varphi_1$, for instance. The full dependence on the scalar fields
can then be 
reconstructed by replacing $\varphi_{ij}^2$ (we drop the subscript 1)
by \eqref{lal}.

The radiative corrections at the one loop level are then obtained by
taking the constant modes as backgound VEVs and
integrating out all the massive modes of the fields. To this end, we
shift $\phi_1\to\phi_1+\varphi$ and at non-zero temperature, we
also introduce a zero mode for the time component of the gauge field
via $A_0\to A_0+\alpha$. The VEVs
contribute to the effective masses for the modes and the one-loop
correction involves the logarithm of the resulting fluctuation
determinants and hence depends on the VEVs in a non-trivial way.
It is important to note to leading order in the coupling in the near
critical region, we can take $\mu_1=R^{-1}$ exactly.

Before we compute these determinants, we must first fix the gauge.
We find it convenient to do this by working in a 
conventional $R_\xi$ gauge of a 
spontaneously broken gauge theory and then to specialize to Feynman
gauge $\xi=1$.\footnote{We shall not be unduly concerned by the usual
  $\xi$ dependence of the 
  effective potential that generally plagues gauge theories. The
  reason is that in the vicinity of the  critical region there is an
  exactly flat 
  direction at tree level and so the VEV can be taken to be non-zero
  whilst remaining on shell. The $\xi$ dependence then drops out as
  one can explicitly find by including it in all subsequent steps.}
We add to the action 
the gauge fixing term
\EQ{
{\cal L}^\text{(gf)}=\frac1{2g^2}\text{Tr}\Big(\nabla_iA^i+\tilde
D_0A^0-i\varphi\phi_1\Big)^2\ .
}
In the above equation and in what follows, 
we leave adjoint action by $\varphi$ as implicit,
{\it i.e.}~$\varphi\phi\equiv[\varphi,\phi]$,
$\varphi^2\phi\equiv[\varphi,[\varphi,\phi]]$, etc.
In addition,
$\tilde D_0=\partial_0+i\alpha$ includes $\alpha$ the zero mode part
of $A_0$ only, and, as for $\varphi$, adjoint action for $\alpha$ is
implied.

Although in the absence of chemical potentials the gauge fixing
removes cross terms between the gauge field 
and the scalars, the presence of a chemical potential and a VEV
$\varphi$ introduces additional ones that are not removed by the gauge
fixing term. The modes $A_0$, $\phi_1$ and 
$\phi_2$ are all coupled together as seen from the expression for the
bosonic part of the action at Gaussian order in quantum fluctuations:
\SP{
&{\cal L}^{(\rm boson)}=
\frac{1}{g^2}\text{Tr}\,\Big[\tfrac{1}{2}A_0(-\tilde
D_0^2-\Delta^{(s)}+\varphi^2) 
A_0 + \tfrac{1}{2}A_i(-\tilde
D_0^2-\Delta^{(v)}+\varphi^2) 
A_i\\
&
+\tfrac{1}{2}\phi_1(-\tilde D_0^2-\Delta^{(s)}+\varphi^2)
\phi_1+\tfrac{1}{2}\phi_2(-\tilde D_0^2-\Delta^{(s)}+\varphi^2)\phi_2\\
&+R^{-1}(A_0\varphi\phi_2-\phi_2\varphi A_0+i 
\phi_1\tilde D_0\phi_2-i\phi_2\tilde
D_0\phi_1)+\tfrac{1}{2}\sum_{a=3}^6 \phi_a(-\tilde 
D_0^2-\Delta^{(s)}+\varphi^2 +R^{-2})\phi_a\\
&+\bar c(-\tilde
D_0^2-\Delta^{(s)}+\varphi^2 )c\Big].
}
Here we have set $\mu_1=R^{-1}$, $\Delta^{(s)}$ and $\Delta^{(v)}$ are
the scalar and vector Laplacians on $S^3$, and we have explicitly
included the ghosts $c,\bar c$ in the Lagrangean. The detailed
properties of these Laplacians and their eigenvalues are summarized in
Appendix A. We need only to note that while $\Delta^{(s)}$ has an
$\ell=0$ mode, the vector Laplacian $\Delta^{(v)}$ on $S^3$ does not
have a zero mode. Furthermore, the vector fluctuations $A_i$ can be
decomposed into the image and kernel of the gradient operator
$\nabla^i$ as $A^i= B^i +C^i$ with  with $\nabla_iB^i=0$ and 
$C^i=\nabla^i f$.

The fermionic fluctuations about the non-zero backgrounds for
$\varphi_1$ and $\varphi_2$ are governed by the Lagrangean,
\EQ{
{\cal L}^{(\rm fermion)}=\Tr\left(i\bar \psi_A
  \left(\Dslash-\bar\mu_A\gamma_0\gamma_5\right)\psi_A - \bar\psi_A\left(
\alpha_{AB}^1\varphi_{1}+i\beta^1_{AB}\gamma_5\varphi_{2}\right)\psi_B\right),
}
where 
\EQ
{\bar \mu_A = R^{-1}(\tfrac{1}{2}, \tfrac{1}{2}, -\tfrac{1}{2}, -\tfrac{1}{2}),
}
using \eqref{ferchemi}. The $\psi^A$ are four component Majorana
fermions.

The one loop contribution to the effective potential is obtained by
integrating out each {\em massive} fluctuation, giving rise to
the corresponding fluctuation determinant. Importantly, the
quadratic fluctuation operators for the $(A_0,\phi_1,\phi_2)$
sector and the fermionic sector are off-diagonal in flavour space. 
For example, using that the eigenvalues of $\Delta^{(s)}$ are given by
$\ell(\ell+2) R^{-2}$, for $\ell =0,1,\ldots$, we can evaluate the
determinant of the fluctuation operator coupling the
$(A_0,\phi_1,\phi_2)$ sector for a given spherical harmonic number,
\SP
{
&{\rm det}\MAT{-\tilde D_0^2+\ell(\ell+2)R^{-2}+\varphi^2 & 0 &
  2R^{-1}
\varphi\\\\
0& -\tilde D_0^2+\ell(\ell+2)R^{-2}+\varphi^2& 2i R^{-1}\tilde D_0\\\\
-2 R^{-1}\varphi & -2i R^{-1}\tilde D_0 & -\tilde
D_0^2+\ell(\ell+2)R^{-2}+\varphi^2}\\\\ 
&=\left[(-\tilde D_0^2 + \ell(\ell+2)R^{-2}+\varphi^2)(-\tilde
  D_0^2+\ell^2 R^{-2}+\varphi^2)
(-\tilde D_0^2 + (\ell+2)^2R^{-2}+\varphi^2)
\right].
}
The zeroes of this expression viewed as a polynomial in $\tilde D_0$
yield precisely the energies of the harmonics.
Note that for this, the simplest situation at hand, these match the
results summarized in Table 1, if we set $\mu_1=R^{-1}, \mu_2=\mu_3=0$.
The table shows that introducing a chemical potential leads to a 
re-organization of the energy levels. In the bosonic sector, when
$\mu_1=R^{-1}$ is turned on, the result 
is three new towers of modes that are shifted in such a way that one
of the new towers is identical to the original $A_0$ tower (which can be
cancelled with the $C_i$ and ghosts) while the other two
new towers are the deformations of the original
$\phi_1$ and $\phi_2$ towers. 
For fermions, notice that the original half-integer graded modes are now
integer graded.

The fluctuation determinant for each species then enters the effective
potential at one loop as
\EQ{
V_1 = {T\over 2\pi^2 R^3}\frac{1}{2}\sum_{\rm species}
\sum_{ij=1}^N\sum_{\ell=\ell_0}^\infty d_\ell^{B (F)} \,\log\det\big[-
\tilde D_0^2+\varepsilon_\ell(\varphi_{ij})^2\big]\ ,
\label{gen}
}
where $d_\ell^{B(F)}$ is the degeneracy of bosonic (fermionic) 
modes with angular momentum
quantum number
$\ell$.
The
integer $\ell_0$ is the lower limit
on the angular momentum quantum number. The quantity
$\varepsilon_\ell$ is the energy of the mode in question. The
degeneracy factors $d_\ell^{B(F)}$ are positive or negative depending upon
the the statistics of the corresponding fields.


In Table 1 we summarize the data associated to each set of modes. 
We emphasize that these energies are strictly only correct when
either $\mu_p=0$ or $\mu_p=R^{-1}$. Where sign
choices exist all the possible combinations must be taken.
\begin{table}
\begin{center}\begin{tabular}{ccccc}\hline\hline
Field & $d_\ell$ & $\varepsilon_\ell$ &  $\ell_0$ \\
\hline
$B_i$ & $2\ell(\ell+2)$ &  $\sqrt{R^{-2}(\ell+1)^2+\varphi^2}$ &1 \\   
$C_i$ & $(\ell+1)^2$ &  $\sqrt{R^{-2}\ell(\ell+2)+\varphi^2}$ &1 \\   
$(c,\bar c)$ & $-2(\ell+1)^2$ & $\sqrt{R^{-2}\ell(\ell+2)+\varphi^2}$ &0 \\ 
$(A_0,\phi_{1,2})_1$ & $(\ell+1)^2 $ &
$\sqrt{R^{-2}\ell(\ell+2)+\varphi^2}$ & 0\\ 
$(A_0,\phi_{1,2})_{2,3}$ & $(\ell+1)^2 $ & $\sqrt{R^{-2}(\ell+1\pm
  R\mu_1)^2+\varphi^2}$ & 0\\ 
$\phi_{3,4}$ & $(\ell+1)^2$  
& $\sqrt{R^{-2}(\ell+1)^2+\varphi^2}\pm\mu_2$ & 0\\
$\phi_{5,6}$ & $(\ell+1)^2$ & 
$\sqrt{R^{-2}(\ell+1)^2+\varphi^2}\pm\mu_3$&  0\\
$\psi^A_\alpha$ & $-\ell(\ell+1)$ &
$\sqrt{R^{-2}(\ell+\tfrac12\pm\tfrac12R\mu_1)^2+\varphi^2}\pm\tfrac{\mu_2}2
\pm\tfrac{\mu_3}2$ & 1\\
\hline\hline
\end{tabular}\end{center}
\caption{\small The fields, their degeneracy and energies as a
  function of the chemical potentials with a 
  nonvanishing VEV for $\phi_1$. The expressions for the mode energies
  are only valid when either $\mu_p=0$ or $\mu_p=R^{-1}$.}
\end{table}

The contributions to the effective action are 
standard expressions in thermal field theory. First of all, the
eigenvalues of $i\partial_0$ are $2\pi n/\beta$, for $n\in{\mathbb Z}$ for
bosons. When acting on fermions, due to antiperiodic boundary conditions
for fermions around the thermal $S^1$, the operator $i\partial_0$ has
eigenvalues $2\pi(n+1/2)/\beta$, for $n\in {\mathbb Z}$.
It is standard practice in thermal field theory to perform a Poisson
resummation over $n$ in such a way that each contribution splits
into a piece that describes the theory at $T=0$ and the non-trivial
``thermal'' part which vanishes as $TR\to 0$. The zero temperatue
piece is the Casimir energy in the presence of background expectation
values for fields. So a typical term in the one loop
potential 
\eqref{gen} can be expressed as
\SP{
&
{1\over{\rm Vol}(S^3)}\sum_{\ell=\ell_0}^\infty d_\ell
\log \,\det(-\tilde D_0^2+\varepsilon_\ell(\varphi)^2)=
\\\\
&{\rm Bosonic}:\qquad 
{1\over {\rm Vol} (S^3)}\frac12\sum_{ij=1}^N\sum_{\ell=\ell_0}^\infty
d_\ell^B\Big\{|\varepsilon_\ell(\varphi_{ij})|- 
\frac1\beta\sum_{n=1}^\infty\frac1n  
e^{-n\beta|\varepsilon_\ell(\varphi_{ij})|}\cos(n\alpha_{ij}\beta)\Big\}\
,\\\\
&{\rm Fermionic}:\qquad
{1\over {\rm Vol} (S^3)}\frac12\sum_{ij=1}^N\sum_{\ell=\ell_0}^\infty d_\ell^F
\Big\{|\varepsilon_\ell(\varphi_{ij})|-
\frac1\beta\sum_{n=1}^\infty\frac{(-1)^n}{n} 
e^{-n\beta|\varepsilon_\ell(\varphi_{ij})|}\cos(n\alpha_{ij}\beta)\Big\}
\label{typical}
}
where we have written the adjoint trace explicitly and defined
$\varphi_{ij}\equiv\varphi_i-\varphi_j$ and $\alpha_{ij}=\alpha_i-\alpha_j$. 

\subsubsection*{ More than one non-zero critical chemical potential}

The analysis with more than one non-zero critical chemical potential
proceeds similarly to the above. The only difference lies
in the mixing matrices for bosonic and fermionic fluctuations and
their resulting eigenvalues. 

When $\mu_1 \simeq \mu_2\simeq R^{-1}$ and $\mu_3=0$, the light scalar
modes are $(\varphi_1,\ldots \varphi_4)$ and a $5\times 5$ mixing
matrix results for the fluctuations in the bosonic sector. The zeroes
of the determinant of this mixing matrix in frequency space correspond
to the mode energies. Similarly one may find the fluctuation energies
for the case of three critical chemical potentials
when all six scalar fluctuations and $A_0$ are coupled. The
results for all the mode energies, both bosonic and fermionic are
listed in Table 1. 



\section{The zero temperature effective potential }

As we take the low temperature limit of \eqref{typical}, the
thermal contributions are exponentially suppressed by the
Boltzmann factors. Then we are only left with a Casimir energy sum
that yields the one loop effective potential
\EQ
{
V_1 (T=0)\,= {1\over{\rm Vol}(S^3)}\,\frac12\,\sum_{\rm Species}\;
\sum_{ij=1}^N\sum_{\ell=\ell_0}^\infty d_\ell^{B(F)} 
|\varepsilon_\ell(\varphi_{ij})|.
}
As is well known, this sum is formally divergent and needs to be
regulated. 

The choice of regulator is a subtle issue especially in the presence
of chemical potentials and there is more than one way to regulate such
sums. We choose to cut off the sums keeping two crucial points in
mind: First, the cutoff will be imposed on the {\em energies} of the
modes rather than the angular momenta $\ell$ so that the 
regulator is general coordinate
invariant.\footnote{ See the footnote (30) 
in \cite{Aharony:2003sx} and Chapter 6 of
  \cite{Birrell:1982ix} for a discussion of these issues.} Secondly,
the energy cutoff function will be chosen to be completely independent
of the chemical potential(s). What this means is that, all mode
sums will be regulated using the cutoff functions of the theory with
$\mu_1=\mu_2=\mu_3=0$. This requirement may be motivated by the
physical observation that introducing a chemical potential is a
deformation of the {\em state} of the theory and not the functional
integral measure. 
 
For each field type, we introduce the regulated expression for the
energy sums
%
\EQ{
\mathscr
E_\text{reg}=\,\frac{1}{2}\,\sum_{ij=1}^N\sum_{\ell=\ell_0}^\infty
d_\ell^{B(F)} 
\,|\varepsilon_\ell(\varphi_{ij})|
f(\varepsilon^{(0)}_\ell/\Lambda).
}
Here $f(x)$ can be thought of as the smooth version of a cutoff
function which is unity for $x<1$ and 
0 for $x>1$ such that $f(0)=1$ while all the derivatives
$f'(0)=f^{\prime\prime}(0)=\cdots=0$. Crucially, the cutoff is on the
energy $\varepsilon_\ell^{(0)}$ 
which is  defined to be the energy of the respective mode with $\mu_p=0$.

We can evaluate the regulated sum by using the Abel-Plana formula
\cite{Mostepanenko:1997sw} appropriate to a function with branch cuts
on the imaginary axis or the left 
half plane
\EQ{
\sum_{n=0}^\infty F(n)=\int_0^\infty
dx\,F(x)+\frac12F(0)-2\int_0^\infty
dx\,\frac{\text{Im}\,F(ix)}{e^{2\pi x}-1}\ .
\label{ap}
}
Let us now turn to the evaluation of each field contribution to the
Casimir energy seriatim.

The first point that is immediately clear is that 
large cancellations occur between the $C_i$, the ghosts $(c,\bar c)$
and $(A_0,\phi_1,\phi_2)_1$ which is basically the $A_0$ field. Indeed, all
the $\ell>0$ contributions cancel between these, leaving a net result
$-\tfrac12|\varphi|$.

The regularized contributions from the $B_i$ and the scalar fields
remain in the bosonic sector. Taking care to use the $\mu_p=0$
energies in the cutoff functions and applying the Abel-Plana formula,
we find that all bosonic modes together induce the one loop potential
\SP{{\mathscr E}_B=&\Lambda^4 R^3 -\frac{1}{2} R \Lambda^2
- R^3\varphi^2\Lambda^2+{1\over 12 R}- {1\over 4}\varphi^2 R
\\\\
&+{1\over 2}\varphi^4 R^3
\log\left({|\varphi| e^{1/4}\over 2\Lambda}\right)+
8\int_{R\varphi}^\infty {x^2 \sqrt{x^2 R^{-2}-\varphi^2}\over 
e^{2\pi x}-1}. 
}
Interestingly, the potentially problematic linear term
$-\tfrac12|\varphi|$ from the ghosts, $A_0$ and the $C_i$, cancels off
in the full sum. 
The result is valid for any number of critical chemical
potentials. The sum over energies for more than one critical
chemical potential reduces to the one for a single chemical potential
due to a simple cancellation of the dependence on $\mu_2$ and
$\mu_3$. This cancellation is obvious from the fluctuation energies of
$\phi_3,\ldots\phi_6$ listed in Table 1. 

For the fermions, again employing a cutoff on the mode energies at
zero chemical potential, the Casimir energy is 
\SP
{{\mathscr E}_F=-&\Lambda^4 R^3 +\frac{1}{2} R \Lambda^2
+R^3\varphi^2\Lambda^2+{5\over 48 R}+ {1\over 4}\varphi^2 R
\\\\
&-{1\over 2}\varphi^4 R^3
\log\left({|\varphi| e^{1/4}\over 2\Lambda}\right)-
8\int_{R\varphi}^\infty {x^2 \sqrt{x^2 R^{-2}-\varphi^2}\over 
e^{2\pi x}-1}.
} 
The total one loop quantum effective potential on $S^3$ is therefore
\EQ
{
{1\over {\rm Vol} (S^3)} \left({\mathscr E}_F+{\mathscr E}_B\right)=
{1\over{\rm Vol}(S^3)}\;{3\over16 R}.
}
This is remarkable since all the dependence on the background VEVs
has completely cancelled out to yield the 
zero point energy of ${\cal N}=4$ theory
on $S^3$. This implies that at least at one loop order,
at the critical chemical potential, there is a complete cancellation
between bosonic and fermionic quantum fluctuations resulting in flat
directions in the 
quantum effective potential.\footnote{It is worth contrasting this
  result with the 
same calculation done at zero temperature and vanishing
chemical potentials. In that case one obtains \cite{Hollowood:2006xb}
a complicated one loop 
effective potential on $S^3$.} 

To summarize the result of the computation presented in this section:
We have found that at the critical value for one, two or
all three chemical potentials of ${\cal N}=4$ theory on $S^3$ at {\em zero 
temperature}, there is no radiatively induced potential for the scalar
fields at one loop
order. This implies, at least at the one loop order, a Coulomb branch
``moduli space'' of complex dimension $N$, $2N$ or $3N$, depending on
whether one, two or three chemical potentials are at the critical
value.

Below we offer 
a natural explanation for the appearance of these flat directions
in the effective potential calculation,
and its existence to all orders in the interacting theory at zero
temperature.

\subsection{Massless BPS states at critical chemical potential}

At zero temperature, chemical potentials for the three $U(1)_R$ charges
deform the Hamiltonian on $S^3$ as,
\EQ
{
\Delta \to\Delta -\sum_{p=1}^3 \mu_p J_p\,.
\label{defham}
}
The ${\cal N}=4$ superconformal algebra \cite{Kinney:2005ej} includes
the following  
anticommutation relation between the superconformal generators
$S_{A\alpha }$ and the supercharges
$Q^{B\beta}$. Noting that  on $S^3$, following the rules of radial
quantization, $\left(Q^{A\alpha}\right)^\dagger= S_{A\alpha}$, we have
\EQ
{
\{Q^{A\alpha\dagger},Q^{B\beta}\}=\delta^B_A\;\delta^\beta_\alpha\;
\Delta+\sum_p\left(J_p
  \,R_p\right)^B_A\;
\delta^\beta_\alpha +\;\delta^{B}_A \,\left({\mathcal
    J}_1\right)^\beta_\alpha\,.
\label{bps}}
A similar relation holds for $\bar Q$ and $\bar S$.
The indices $A, B=1\ldots 4$ are $SU(4)_R$ indices
and $R_p$ ($p=1,2,3$) are the three $U(1)_R$ Cartan generators in the
fundamental representation of the $SU(4)_R$ algebra; ${\mathcal
  J}_1$ is a generator of an $SU(2)\subset SO(4)$ isometry group of
$S^3$. 

For a subset of the supercharges, 
the right hand side of the above relation 
vanishes on BPS states of the ${\cal N}=4$ theory. 
Multi-trace operators with 
R charge $(J_1, 0, 0)$ and with $(R\Delta)-J_1=0$ are
$\tfrac{1}{2}$ BPS states which are annihilated by one half of the
chiral supercharges in \eqref{bps}. Similarly, 
states satisfying $(R\Delta)- J_1-J_2=0$ and $(R\Delta)-J_1-J_2-J_3=0$ correspond to 
${\tfrac{1}{4}}^{\rm th}$ BPS and ${\tfrac{1}{8}}^{\rm th}$ BPS states
respectively. 

The operators $\Delta-J_1/R$; $\;\Delta- (J_1+J_2)/R$ and
$\Delta-(J_1+J_2+J_3)/R$ are of course the Hamiltonians of the theory
with critical chemical potentials \eqref{defham}.
It is thus clear that the ground states of the Hamiltonian
with critical chemical potentials
 are infinite sets of one half, one
quarter and one-eighth BPS states of the ${\cal N}=4$ theory, depending
on whether we have one, two or three critical chemical
potentials. Importantly, in each case there is an infinite degeneracy
of ground states, all of which have zero energy, following from
\eqref{bps}. 

Choosing an appropriate ${\cal N}=1$ subalgebra of the ${\cal N}=4$
superalgebra, the chiral ring is the set of holomorphic gauge invariant
operators made up of polynomials of $\Phi_1=\phi_1+i\phi_2$,
$\Phi_2=\phi_3+i\phi_4$ and $\Phi_3=\phi_5+i\phi_6$, and the chiral
gauge field strength $W_\alpha$, modulo F-term relations (on
${\mathbb R}^4$). These are the $\tfrac {1}{8}$ BPS states. On
${\mathbb R}^4$, the chiral primary operators at a generic point on 
the Coulomb branch of the theory, are the 
the same as the operators of the chiral ring described above. The F-term
conditions ensure that the fields may be simultaneously diagonalized
and the chiral operators are distinct polynomials of the 3$N$ bosonic
and $2N$ fermionic diagonal eigenvalues, invariant under the
permutation group $S_N$. 
On $S^3$,
these operators are built out of the $s$-waves of the diagonal
elements of the three complex scalars
and two polarizations of the $\ell=1$ harmonic of a fermion 
\cite{Berenstein:2005aa}.

Similarly, the half BPS states correspond to $S_N$ invariant 
polynomials involving the diagonal elements of the 
$s$-wave of a single bosonic holomorphic field, namely
$\varphi_1+i\varphi_2$. The quarter BPS states are generated by
diagonal elements of $s$-waves of the 
holomorphic bosonic operators $\varphi_1+i\varphi_2$ and
$\varphi_3+i\varphi_4$. 

The dimensions of the moduli spaces encountered at the different
numbers of critical chemical potentials, and the 
corresponding counting of zero
modes is therefore perfectly consistent with the interpretation that
at critical chemical potential BPS states of the ${\cal N}=4$ theory
become light. Each point on the resulting moduli space represents a
coherent superposition of the light BPS states.

\section{Low temperatures and a metastable plasma phase}

\begin{figure}[h]
\begin{center}
\epsfig{file=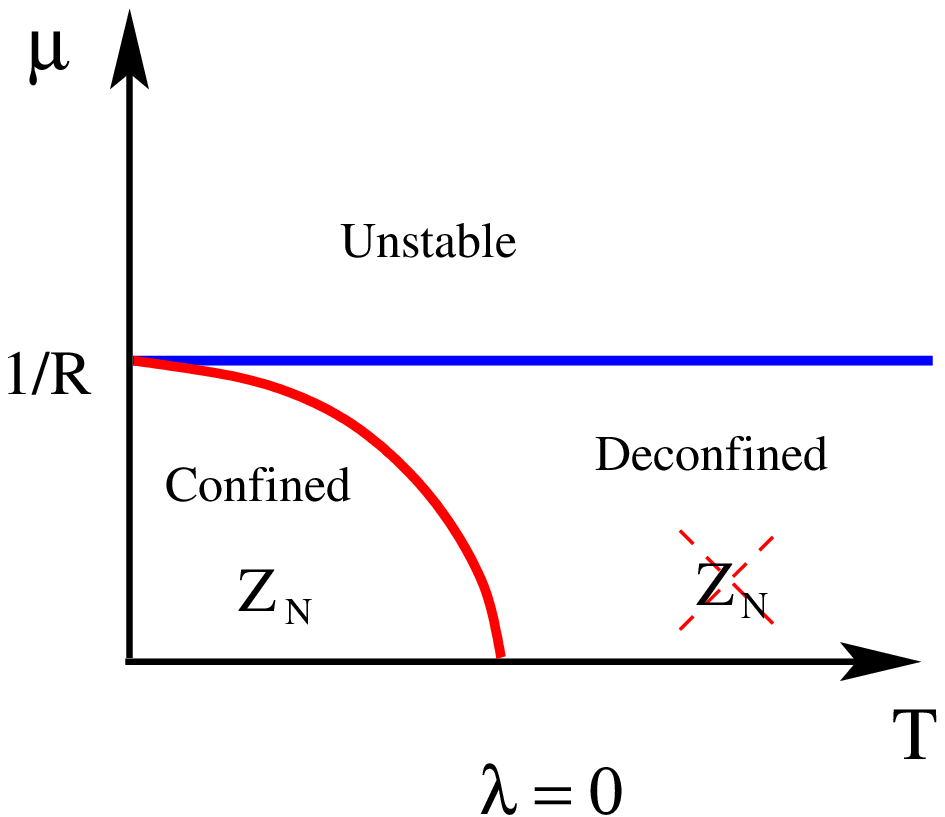,width=2.5in}
\end{center}
\noindent \small {\bf Figure 1:} The $\lambda=0$ phase diagram as a function of
  temperature and chemical potential, found by \cite{Yamada:2006rx}.
\label{phase1}
\end{figure}

To summarize the story thus far, the phase diagram of the free
theory was studied in  \cite{Yamada:2006rx} 
at all temperatures and chemical potentials. At infinite $N$, the authors of 
\cite{Yamada:2006rx}  established the existence 
of a first order Hagedorn/deconfinement  transition for any fixed
$\mu_p$ as the temperature is increased (see Figure 1.). For $\mu_p >
R^{-1}$, the free theory ceases to have a well-defined 
ground state in the grand canonical ensemble.

In the vicinity of $\mu_p\simeq R^{-1}$, we have shown that light 
scalars appear and remain light in the interacting theory at
$T=0$. Furthermore, we see flat directions associated to these light modes.
For a small non-zero temperature $TR\ll 1$, we expect the scalar modes 
$(\varphi_{2p-1},\varphi_{2p})$ to continue to
be light degrees of freedom in the near critical region. In addition
there will be another set of zero modes in the theory, namely
the $\alpha_i$, the diagonal elements of the Polyakov loop matrix.

\subsection{Low temperatures and
$\mathbf{\mu_1=R^{-1}, \mu_2=\mu_3=0}$}
At the critical chemical potential, and at zero temperature,
the effective potential is flat including quantum effects. 
There are no non-analyticities in the effective potential 
near the origin, even when off-diagonal
modes become light. This is due to an exact cancellation between the
Vandermonde measure factor (equivalently the ghosts and $A_0$) and the
contribution from the new scalar zero mode which appears at the critical
chemical potential. 

Our strategy will be to stay at $\mu_1={R}^{-1},\mu_2=\mu_3=0$ and
switch on a small non zero temperature $TR\ll1$.  The finite
temperature effective potential 
at this critical point is
\SP
{
&V_0+V_1=\\
&\sum_{ij=1}^N {1\over {\rm Vol}(S^3)}\left[{3\over 16R}
- 8 T \sum_{k=1}^\infty{1\over 2 k-1}\; \cos\left( (2 k
  -1){\alpha_{ij}\over T}\right)\; 
\sum_{\ell=1}^\infty \ell^2 \; 
e^{-{(2k-1)\over T}\sqrt{\ell^2 R^{-2}+\varphi_{ij}^2}}\right].
}
The effective potential above is a function of the light fields 
\EQ
{
\varphi_{ij}^2=(\varphi_{1i}-\varphi_{1j})^2+(\varphi_{2i}-\varphi_{2j})^2.
}

A notable feature of this expression is that there are
no non-analytic terms near the origin where we expect off-diagonal
modes of $\phi_1$ and $\phi_2$ to become light. 
The exact cancellation between the scalar zero mode and
measure factors persists at non-zero temperature. 
This is related to the fact that with one critical
chemical potential, 
the ground states at zero temperature are 
parametrized by multi-trace gauge-invariant combinations of the 
$s$-wave of a {\em single} holomorphic field, namely
$\varphi_1+i\varphi_2$. All other modes are heavy. At the origin,
therefore, we expect that even if off-diagonal modes of the field
appear to become light at non-zero temperature, 
we may gauge rotate them away completely. 
(By the same logic, we expect not to be able to do this for the
$\tfrac{1}{4}^{\rm th}$ and $\tfrac{1}{8}^{\rm th}$ BPS states, since
they involve multi-trace combinations of more than one holomorphic
field.)

The radiative corrections vanish exponentially at large field
amplitudes $|\varphi_{ij}|R\gg 1$. This is actually a generic,
robust feature expected of the effective potential of ${\cal N}=4$
theory on $S^3$. In the limit of large field amplitudes $|{\varphi_{ij}| R}
\gg 1$, all off-diagonal
modes are proportionately heavier and therefore should decouple
from the theory. That this happens unambiguously is linked to the UV
finiteness of the ${\cal N}=4$ theory. Furthermore, as already noted
above, the putative
scalar zero energy mode (near $\varphi_{ij}^2\approx0$) 
has cancelled against the ghosts.

Now we need to determine the minimum of this joint potential.
It is clear that the $\alpha_i$ experience a purely attractive 
pairwise potential near $\varphi_{ij}=0$. This is because the
repulsive Vandermonde determinants have been eliminated through
cancellations with the zero mode. Hence at finite temperature, at the
critical chemical potential,
\EQ
{
\alpha_1=\alpha_2=\ldots=\alpha_N;
}
which means that the large $N$ theory should be thought of as being
deconfined and the large $N$ distribution of the $\alpha_i$ is a
delta function.
With $\alpha_{ij}=0$, it is clear that at
$\varphi_{ij}=0$, the scalars have a mass squared which is strictly
positive and the origin is the stable vacuum of the theory.

Expanding the potential for $R^2\varphi^2_{ij}\ll1$, and $TR\ll 1$,
we see that
thermal effects contribute a small mass to the scalars
\EQ
{V_0+V_1\;\approx\;
{N^2\over 2\pi^2 R^4}\left({3 \over 16}
-8 T R \;e^{-{1\over TR}}
+ \;2\;R^2\; e^{-{1\over TR}}
{1\over N}\sum_{i=1}^N (\varphi_{1i}^2 
+\varphi_{2i}^2 )
\right ).
}
Interestingly, a small temperature $TR\ll1$ has lowered the vacuum energy
at the origin to ${3/16R  - 8 T e^{-1/TR}}$ and 
the light scalars $\varphi_1$ and $\varphi_2$ have acquired 
an exponentially
small positive mass.

\subsection*{Metastable plasma for $\mathbf{\mu_1\gtrsim
  R^{-1},\mu_2=\mu_3=0}$}

We now want to argue that since a small temperature makes the light
scalars massive at $\mu_1=R^{-1}$, the origin will continue to be a
locally stable vacuum if we raise $\mu_1$ by a sufficiently small amount 
above its critical
value. This is of course only possible if the thermal mass can beat
the instability induced by the chemical potential. Crucially, we
also require that, for our analysis to apply
\EQ
{
\mu_1-R^{-1}\lesssim {\cal O} (\lambda),
}
which allows us to set $\mu_1= R^{-1}$ in the one loop calculation,
the error in doing so appearing at one higher order in perturbation theory.

With this choice of parameters, we have
\EQ
{
V_0+V_1=\sum_{ij=1}^{N}\left({1\over
    4\lambda}(R^{-2}-\mu_1^2)\varphi_{ij}^2
+{1\over {\rm Vol}(S^3)}\left[{3\over 16 R}-8T\; e^{-{1\over
        TR}\sqrt{1+ R^2\varphi_{ij}^2}}\right]\right).
 }
Since the quantum correction vanishes exponentially
at large field amplitudes, the potential will revert to its classical
runaway behaviour in these regions whenever $\mu_1 > R^{-1}$. Close to
the origin, however, the situation is different. For the following range
of values of $\mu_1$
\EQ
{
0<\mu_1-R^{-1}< \tfrac{4\lambda} {\pi^2 R} \;e^{-{1\over  T R}},
}
the theory has a metastable vacuum at the origin. The behaviour of the
effective potential is shown in Figure 2.

\begin{figure}[h]
\begin{center}
\epsfig{file=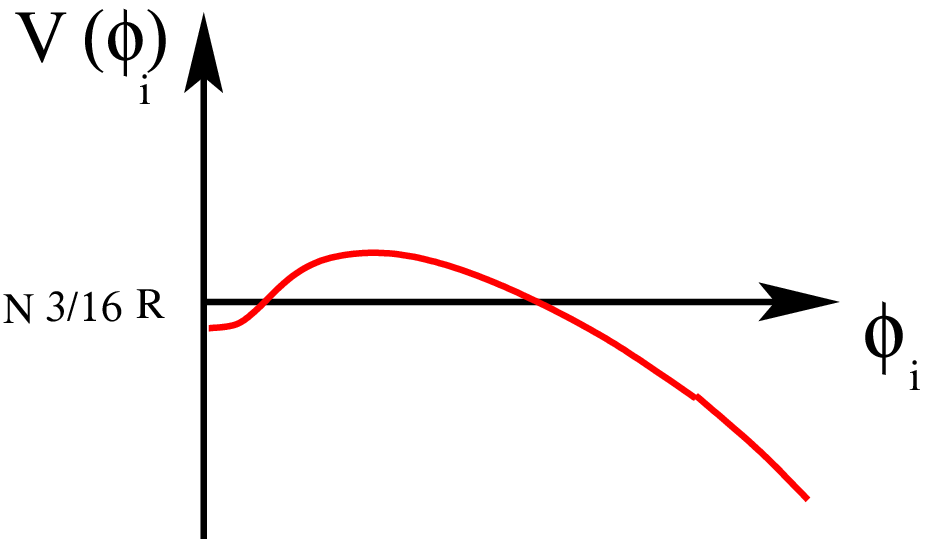,width=2.5in}
\end{center}
\noindent \small {\bf Figure 2:} The low temperature $TR\ll1$ 
effective potential for one
diagonal mode of the light scalar with 
$0<\mu_1 -R^{-1}\leq (
4\lambda /\pi^2 R)\exp(-\frac{1}{TR})$ and $\mu_2=\mu_3=0$. 
\label{phase2}
\end{figure}

It is reasonable to assume that this metastable phase is the
continuation to low temperatures, of the high temperature metastability
found in \cite{Yamada:2006rx}. As in the high temperature situation,
the lifetime of the state is determined by the probability for one
diagonal mode, say $\varphi_{1i}$, to make its way out into the
unstable region. This can happen either by thermal activation over the
barrier, or by tunnelling through the barrier. The probability for
thermal activation can be estimated by computing the Boltzmann
suppression factor. At low temperatures , the height of the barrier is
exponentially suppressed and $\propto \exp(-\tfrac{1}{TR})$, so that
the associated probability for thermally exciting 
one eigenvalue goes as $\exp\left(- N e^{-1/TR}\right)$. Estimating
the barrier penetration probability is more complicated due to the
form of the effective potential, but the dependence on $N$ is
obvious, due to the scaling of the action with $N$.  The
lifetime of the metastable plasma diverges in the strict limit $N\to
\infty$, as was also found for the high temperature metastable phase
in \cite{Yamada:2006rx}.
 
\subsection{Two and three non-zero critical chemical potentials}

We have seen that with two and three critical $\mu_p$, the theory at 
$T=0$ has flat directions which are not lifted by quantum
corrections. 

At finite temperature however, the one-loop effective potential for
the diagonal modes with  
two and three critical chemical potentials 
displays some qualitatively different features compared to the case with
one chemical potential. The main difference is due to the appearance
of additional zero modes for small $\varphi_{ij}$. The
potential for the diagonal modes exhibits non-analytic behaviour near
the origin. We may understand these non-analyticities as being 
due to the off-diagonal modes of the $s$-wave components of
the light holomorphic fields getting excited near the origin, at finite
temperature. 

At zero temperature and critcial $\mu_p$, the light Coulomb branch states
consist of $\tfrac{1}{4}^{\rm th}$ and $\tfrac{1}{8}^{\rm th}$ BPS
states. However, close to $\varphi_{ij}=0$ at finite temperature, there are
also light off-diagonal excitations which are non-BPS. Near the
origin, therefore, it is more appropriate to look at the perturbative
result for the effective potential for the $s$-wave of the
full matrix fields obtained by integrating out higher harmonics. As we
will see below the only effect of this is to provide a small positive thermal
mass.
\\\\
\underline{\it Two critical chemical potentials}

The low temperature analysis $(TR\ll1)$ 
for two near critical chemical potentials
yields the following one loop correction to the effective potential for
the diagonal modes,
\SP
{
V_1=&
\sum_{ij=1}^N\;{1\over {\rm Vol}(S^3)}\;\Big[{3\over 16R}
-{T\over 2} \;\sum_{n=1}^\infty\;{1\over n}\;\cos\left(n\alpha_{ij}\over
  T\right)
e^{-{n\over T R}\sqrt{1 +\varphi_{ij}^2 R^2}+{n\over TR}}\\\\
&-2\;T
  e^{-{1\over TR}\sqrt{1+\varphi_{ij}^2 R^2}+{1\over
      2TR}}\;\cos\left({\alpha_{ij}\over  
    T}\right)+
{\cal
      O}(e^{-{1\over TR}})\Big].\label{off12}
}
Here $\varphi_{ij}^2= \sum_{a=1}^4(\varphi_{ai}-\varphi_{aj})^2$. Once
again, near $\varphi_{ij}=0$, the $\alpha_i$ experience a purely
attractive potential so that $\alpha_i=\alpha_j$ and the theory is in
the deconfined phase. This expression for the effective
potential exhibits non-analytic behaviour at the origin. With
$\alpha_{ij}=0$, we have
\SP{
V_1=&
\sum_{ij=1}^N\;{1\over {\rm Vol}(S^3)}\;\Big[{3\over 16R}+
{T\over 2}\;\log
\left(1-e^{-{1\over T R}\sqrt{1 +\varphi_{ij}^2 R^2}+{1\over TR}}\right)\\\\
&-2T\;
  e^{-{1\over T R}\sqrt{1+\varphi_{ij}^2R^2}+{1\over 2TR}}+\ldots\Big].
}
Near the origin, the logarithm provides an infinite {\em attractive}
potential and is the result of integrating out off-diagonal elements
of the $\ell=0$ harmonic
of $\varphi_3+i\varphi_4$. The second term which is ${\cal
  O}(e^{-1/2RT})$, originates from integrating out the next lightest
state, namely the fermion with mass ${1/2R}$. Expanding this contribution
about $\phi_{ij}=0$,
\EQ
{
-{1\over {\rm Vol}(S^3)}\;2\;T
  e^{-{1\over T R}\sqrt{1+\varphi_{ij}^2 R^2} +
    {1\over2RT}}\approx{1\over 2\pi^2 
    R^3}\left(-2T\; e^{-{1\over 2RT}}+ R\;e^{-{1\over 2RT}}\;\varphi_{ij}^2
\right),
}
\\
we see a thermal mass for the diagonal modes. By gauge invariance at
the origin, the off-diagonal fluctuations will also obtain the same
thermal mass. The resulting picture is therefore rather similar to
what we have seen earlier. Instead of integrating out light
off-diagonal modes near the origin, we may simply compute the thermal
mass of the lightest fields, namely $\varphi_1+i\varphi_2$ and
$\varphi_3+i\varphi_4$ by computing the usual Feynman one-loop 
self-energy graphs at finite temperature. At
quadratic order, the
thermal effective potential for these fields close to the origin will have
the form
\EQ
{
V= {N^2\over 2\pi^2
    R^3}\left({3\over 16 R}-2T\; e^{-{1\over 2RT}}+ 
2\;R\;e^{-{1\over 2RT}}\;\sum_{p=1}^4{1\over
  N}\Tr(\varphi_a^2)\;+\;{\rm quartic} 
\right).
}
The higher order interactions are the terms responsible for inducing
non-analyticities in the effective potential for the diagonal modes
when off-diagonal fluctuations are (wrongly) integrated out close to
the origin of field space.
 
Hence, a finite temperature introduces a small dip in the potential
energy at the origin, which then asymptotes to the constant value of
$3/(16R)$ at large field amplitudes in accordance with \eqref{off12}.
\\\\
\underline{\it Three critical chemical potentials}
 
The situation with three critical chemical potentials is similar to
that with two critical chemical potentials. There is an additional
light scalar as well as a fermion zero mode.
The potential for the light diagonal modes is 
\SP
{
V_1=
&
\sum_{ij=1}^N\Big(
{1\over {\rm Vol}(S^3)}\Big[{3\over 16R}
-2\;T \sum_{k=1}^\infty{1\over 2k-1}\cos\left((2k-1){\alpha_{ij}\over
  T}\right)
e^{-{(2k-1)\over
    RT}\left[\sqrt{1+\varphi_{ij}^2R^2}-1\right]}
\\\\
&
-4\;T\;\left(e^{-{1\over TR}\sqrt{1+\varphi_{ij}^2R^2}}
+2\;e^{-{1\over TR}\left[\sqrt{1+\varphi_{ij}^2R^2}-1\right]}
\right)\cos\left({\alpha_{ij}\over T}\right)+\ldots\Big]\Big).
}
As before the $\alpha_i$ experience an attractive potential resulting
in a delta-function distribution at large $N$ for these fields. There
is then a logarithmic attractive potential between the light scalar
diagonal modes. It is therefore more appropriate to keep the
all $s$-wave fluctuations of the matrices
$\varphi_1,\varphi_2,\ldots\varphi_6$ while
integrating out all higher partial waves on $S^3$. This leads to a
temperature induced positive curvature in the effective
potential near the origin
\EQ 
{V\approx {N^2\over 2\pi^2 R^3}\left({3\over 16 R}- 12\;T\;
    e^{-{1\over TR}}+ 8\;e^{-{1\over TR}}\sum_{a=1}^6{1\over N}
\Tr \varphi_a^2 + \ldots\right).
}
At large field amplitudes the effective potential approaches a
constant, given by the Casimir energy of the ${\cal N}=4$ theory on
the sphere.

\subsection*{Metastable phase}

Since we have established a positive curvature for the effective
potential at the origin, the argument for the existence of a
metastable plasma phase above critical chemical potential proceeds
exactly as in the situation with one chemical potential.

For two equal chemical potentials, the metastable phase exists provided
the larger of the two chemical potentials exists in the range
\EQ
{
0<\mu-R^{-1}< {2\lambda\over\pi^2 R} \;e^{-{1\over 2TR}}.
}

For three chemical potentials, the metastable plasma phase exists if
the largest of the chemical potentials satisfies
\EQ
{
0<\mu-R^{-1}< {8\lambda\over\pi^2 R} \;e^{-{1\over TR}}.
}
In both cases, the lifetime of this phase is determined by the
probability for one diagonal mode to make it out into the unstable
region. This probability is exponentially suppressed in the large $N$ limit.

\begin{figure}[h]
\begin{center}
\epsfig{file=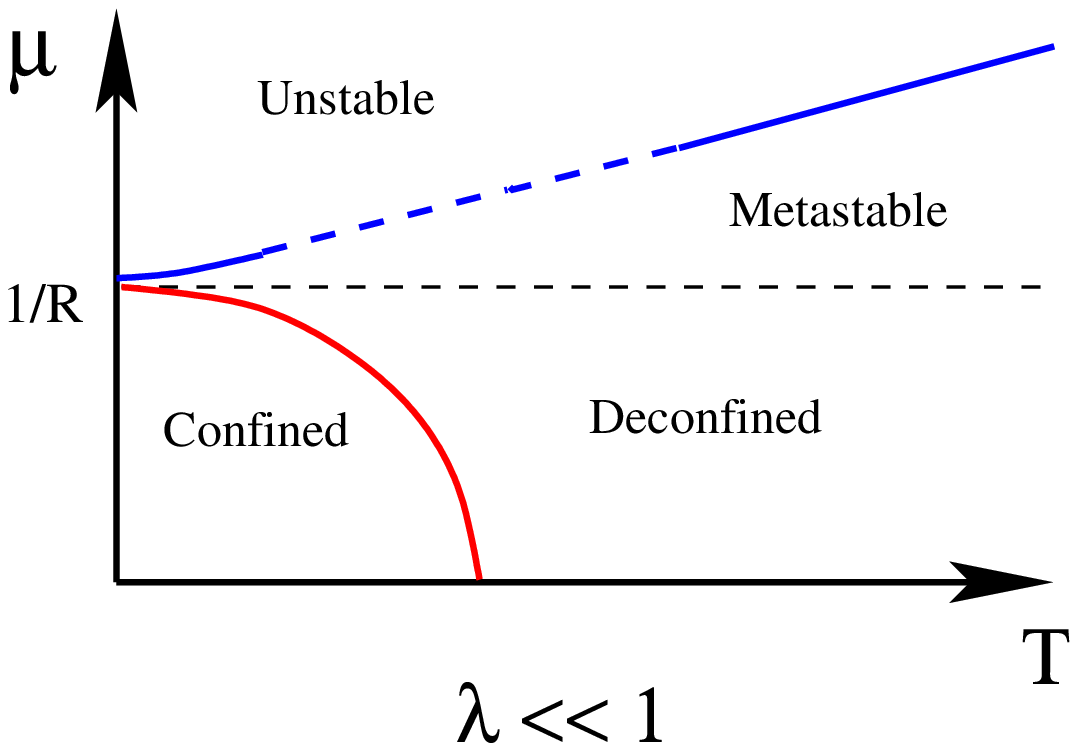,width=3in}
\end{center}
\noindent \small {\bf Figure 3:} The known features of the weak coupling 
phase diagram of ${\cal N}=4$ theory with a chemical potential. 
\end{figure}

\subsection{Metastable plasma at high temperatures}

The picture that we have found above at low temperatures and near the
critical chemical potential, nicely complements the results
of \cite{Yamada:2006rx}. In \cite{Yamada:2006rx}, a careful high
temperature analysis of the theory with chemical
potentials on $S^3$ was performed. The main observations therein maybe
summarized as follows. At high temperatures $TR\gg 1$, the theory
behaves much like the theory on flat space. One may then derive the
``electric'' effective theory by integrating out all non-static 
fluctuations and all fluctuations on length scales of order
$1/T$. This theory is valid on length scales of the order of the Debye
or electric screening scale $\sim (\sqrt \lambda
T)^{-1}$. Specifically this means that the $S^3$ radius $R \sim (\sqrt \lambda
T)^{-1}$ and is parametrically smaller than the nonperturbative
magnetic screening scale. The electric effective theory has an
effective mass for the scalar fields, given by
\EQ
{
m_p^2=R^{-2} -\mu_p^2+\lambda T^2;\qquad {p=1,2,3},
}
where $\lambda T^2$ is the thermal or Debye mass for the scalars. This
means that the effective potential for scalars will have a positive
curvature at the origin, if the largest of the three
chemical potentials satisfies
\EQ
{
\mu_p  < \sqrt{R^{-2}+\lambda T^2}.
}
At large field amplitudes, perturbative quantum corrections are
expected to vanish due to large masses for the heavy states which are
integrated out. In this regime the effective potential will
revert to its classical behaviour. If the largest chemical potential
exceeds $1/R$, then the potential has runaway behviour for large
field amplitudes. Thus we are left with a metastable vacuum at the
origin at high temperatures. The lifetime of this vacuum
diverges
exponentially at large $N$ and as in the cases encountered in this
paper, it is determined by the probability 
for one eigenvalue to make
it out to the unstable region by thermal activation or barrier
penetration. At high temperatures $TR \simeq {1/\sqrt{\lambda}}$ 
this probablity is $\propto \exp(-N
(TR)^3)\sim\exp(-{N\lambda^{-3/2}})$ which vanishes exponentially in
the strict large $N$ limit.

Putting together  the high $T$ results of \cite{Yamada:2006rx} and the
low $T$ results in this article, we obtain a 
phase diagram for the theory at weak coupling, summarized in
Figure 3. (It is, of course, not known as of now whether the first
order line seen in the $\lambda=0$ theory, persists in the interacting theory.)
As we discuss below, this weak coupling phase diagram bears a remarkable
resemblance  to the one at strong coupling (Figure 4). However, there also appear to be
certain intriguing and unexplained differences which we describe in
some detail below.

\begin{figure}[h]
\begin{center}
\epsfig{file=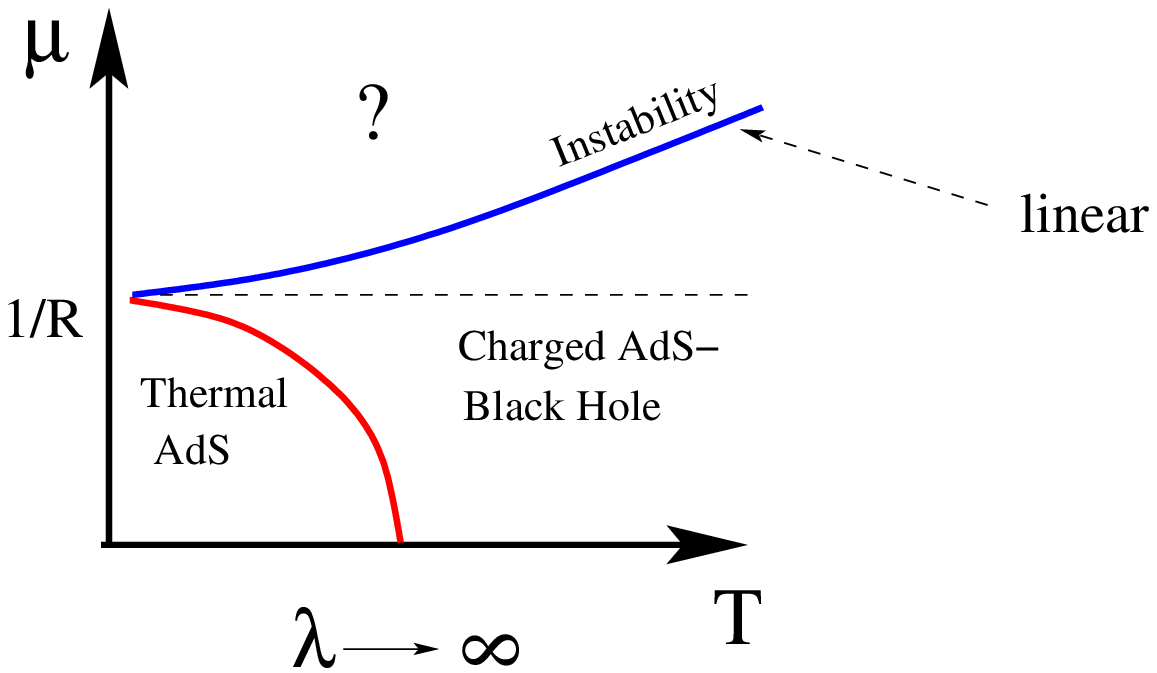,width=3.5in}
\end{center}
\noindent \small {\bf Figure 4:} The
phase diagram of ${\cal N}=4$ theory with chemical potentials at
strong coupling and infinite $N$. The figure corresponds to a generic
pattern of chemical potentials, or angular velocities along the
internal $S^5$ of the gravity dual. It has been argued
\cite{Yamada:2008em} that the region above $\mu =1/R$ and below the
instability line is actually metastable in much the same way as the
field theory at weak coupling.  
\end{figure}

\section{Comparison with strong coupling}

In the strong coupling limit $\lambda\to \infty$, 
the $SU(N)$, ${\cal N}=4$ theory at infinite $N$, 
with R symmetry chemical potentials is dual
to Type IIB supergravity in $AdS_5\times S^5$ with three independent 
angular motions along the internal $S^5$ directions. Working in global
coordinates, there 
are two possible saddle point configurations
that contribute to the semiclassical gravity 
partition function. One of these is the ``spinning'' $AdS_5\times S^5$ geometry
while the other  
is a  charged black hole in AdS.
In the case where all three rotation parameters are equal 
(corresponding to three equal chemical potentials), the gravity dual
reduces to Einstein-Maxwell theory on $AdS_5$ space. The system
undergoes a first order Hawking-Page transition, as the temperature is
increased for fixed chemical potential, from spinning thermal AdS
space to the
Reissner-Nordstrom-AdS black hole \cite{Chamblin:1999tk}. A similar
phenomenon occurs for more general rotation parameters wherein the dual
geometry is described by charged black hole solutions in 5D ${\cal N}=8$
supergravity \cite{Behrndt:1998jd, Cvetic:1999ne}. 

As is apparent from the strong coupling phase diagram in Figure 4,
$\mu=R^{-1}$ is not a point of 
instability, except possibly at $T=0$. Indeed, at $T=0$, setting the
chemical potential equal to $R^{-1}$ corresponds to 
the internal $S^5$ rotating at the speed of light. Crossing this limit
triggers an instability due to the time direction
becoming spacelike near the origin of the space
\cite{Hawking:1999dp}. 

When the theory is in the charged black hole phase
(deconfined plasma), increasing the chemical potential beyond $1/R$ 
does not lead to any local instabilities. Local instabilities are only
triggered at much higher values of the chemical potential
\cite{Cvetic:1999ne, Gubser:2000mm}. The fate of the theory
above this instability line remains unknown.

Figures 3 and 4 illustrate the remarkable qualitative similarities
between the weakly coupled (one loop) field theory on the one hand and
its strongly coupled limit on the other. In fact it has recently been argued
\cite{Yamada:2008em} that the metastable plasma phase seen at
weak coupling has a strong coupling analogue and that the charged 
AdS black holes are metastable for $\mu>R^{-1}$ and below the
instability line. Viewing the charged black holes as near horizon
limits of rotating brane configurations, the system is found to be
metastable to one of the branes splitting from the system of $N$
coincident rotating branes. This is precisely the picture expected from weak
coupling arguments where the locally stable vacuum can decay by one of
the scalar eigenvalues tunnelling out into the unstable region. The
qualitative physical resemblance between weakly coupled gauge theory
and dual gravity at strong coupling lends further support to the idea
that some properties of gravity may be encoded in and extracted from 
weakly coupled gauge degrees of freedom.

However, while pointing out the similarities between the weak and
strong coupling phase diagrams, we should also emphasize the
qualitative differences. These qualitative differences are the outcome
of our analysis at low temperatures in the near critical region, and
continue to pose intriguing questions. Reinterpreting the phase
diagram of \cite{Cvetic:1999ne} in terms of grand canonical variables,
$\mu$ and $T$, it has been pointed out in 
\cite{Yamada:2006rx, Yamada:2007gb} that the phase diagram at strong
coupling has particularly interesting features at low temperatures and
near the critical chemical potential. Specifically, only when all
three chemical potentials are equal $\mu_1=\mu_2=\mu_3=\mu$, the black
hole instability line meets the first order transition line at $T=0$
and $\mu=1/R$. At this point an extremal black hole solution with zero
horizon radius becomes preferred over pure AdS. When the chemical
potentials are unequal however, e.g. $\mu_2=\mu_3=0$, 
one finds that the black hole instability line and the first order
line meet at $\mu_1=1/R$ with $TR=1/\pi$. At this point the charged
black hole horizon shrinks to zero size and the actions of thermal AdS 
and charged black hole coincide. It is not known what happens to the
gravity dual below this temperature. The picture we have found in
weakly coupled field theory is quite different, since the instability
line meets the $\mu=1/R$ line precisely at $TR=0$ in all cases. In all
cases, at weak coupling, the instability line approaches the $\mu=1/R$
line exponentially at low temperatures.

\section{Discussion and future questions}

In this paper we have investigated a corner of the phase diagram of
${\cal N} =4$ theory with R symmetry chemical potentials on $S^3$. The region
of interest corresponds to chemical potentials close to their critical
values, beyond which the theory is known not to have a ground
state. By explicitly computing an effective potential for the light
degrees of freedom in this region, we showed that at the critical
values for the chemical potentials and at zero temperature, 
flat directions open up. In particular, we find moduli spaces
parametrized by $N$, $2N$ and $3N$ bosonic degrees of freedom for one,
two and three critical chemical potentials respectively. These
correspond to $\tfrac{1}{2}$ BPS, $\tfrac{1}{4}^{\rm th}$ BPS and
$\tfrac{1}{8}^{\rm th}$ BPS sates of the ${\cal N}=4$ theory,
respectively becoming light at critical values for different numbers of
chemical potentials. The counting of zero energy modes at the origin
of the moduli spaces is consistent with this interpretation.

Using the above picture of an effective potential, we calculate the
same at small temperatures at critical values of the chemical
potentials. The small positive thermal masses for the light degrees of
freedom allow us to move away from critical chemical potentials, and
indeed to exceed the critical values, resulting in metastable
ground states. In all these situations, the theory is found to be in a
deconfined phase with the eigenvalues of the Polyakov loop collapsing
to a point on the circle. These metastable plasma
phases are the extension to low temperatures, of the metastable states
at high temperature discovered  in \cite{Yamada:2006rx}. Putting
together the results at low and high temperatures, the resulting phase
diagram of the theory at weak coupling is summarized in Figure 3. The
remarkable resemblance to the 
strong coupling phase diagram in Figure 4 is clear. 
However, there are crucial differences as well. The main difference to
emerge from our analysis is that the width of the metastable region
goes to zero only at zero temperature (exponentially). This means that
the instability line meets the first order deconfinement 
transition line at $\mu
=1/R$ and $T=0$ for all patterns of chemical potentials. The strong
coupling picture is quite different. The only situation where the
first order Hawking-Page line meets the black hole instability line
at $T=0$, is when all three chemical potentials are equal. In all
other cases, the instability line and the Hawking-Page line meet at
a finite temperature. Despite these differences 
it is quite remarkable that the
metastability discovered first in weakly coupled field theory, does
appear to have a strong coupling analogue \cite{Yamada:2008em}.

There are several related questions to 
pursue, along the lines of those addressed in this work. One obvious
generalization of our results would be to the $\beta$-deformed theory 
which has three global $U(1)$ symmetries. The free theory will have
the same phase structure as the ${\cal N}=4$ theory. However at
non-zero weak coupling, the theories will differ. Due to lower
supersymmetries, we expect that turning on different numbers of
critical chemical potentials may result in scenarios quite different
from the ${\cal N}=4$ case. Particulary interesting is the question
of the existence of a stable Bose-Einstein condensed state at weak
coupling. It may be possible to look for such a phase by turning on,
say more than one critical chemical potential, followed by a small 
temperature. Looking for such an exotic phase 
in the dual gravity setup would be extremely interesting.

In \cite{Harmark:2006di}, it was pointed out that in the limit, $T\rightarrow
0$, $\mu \rightarrow {R^{-1}}$ with ${T \over R^{-1} - \mu}$ fixed,
the theory reduces to various quantum mechanical sectors,  for e.g.,
the ferromagnetic Heisenberg spin chain. It is interesting to take
this limit in our calculation of the grand canonical partition
function and make contact with the results in recent studies of
integrability in planar ${\cal N}=4$ SYM.   

{\bf Acknowledgements:} We have benefitted from discussions with  
D. Berenstein, N. Dorey, S. Hartnoll, S. Ross, D. Son, 
and L. Yaffe. We thank them for their insights and comments. 
TJH, SPK and AN would like to thank the Isaac Newton Institute, Cambridge
for hospitality during the `` Strong Fields, Integrability and Strings''
workshop, wherein this project received fresh impetus.
SPK would like to thank the KITP, Santa Barbara for hospitality during
the workshop on `` Nonequilibrium dynamics in Particle Physics and
Cosmology'' in course of which 
this project was completed. This research was supported in part by the
National Science Foundation under Grant no. PHY05-51164.

\appendix

\section{Scalar and Vector Laplacians on $S^3$}
Below we review some relevant aspects of 
the spherical harmonic decomposition of fields on $S^3$.
To begin with, consider the kinetic terms for the gauge field:
\EQ{
{\cal L}^\text{(gauge)}=\text{Tr}\,\tfrac14F_{\mu\nu}F^{\mu\nu}=
\int d^4x\,\sqrt{\det\,g}\text{Tr}\big(-\tfrac12
A_\mu(\tilde D_0^2+\Delta)A^\mu-
\tfrac12(\tilde D_0A_0+\nabla_i A^i)^2\big) .
\label{vgg}
}
The Laplacian $\Delta=\nabla_i\nabla^i$
on $\S^3$ depends on the tensorial nature of
the object on which it acts. For example, on the vector component
\EQ{
\Delta A^i=\nabla_j\nabla^jA^i-R^i{}_jA^j\ ,
}
where $R_{ij}$ is the Ricci tensor of $\S^3$. For the 
component $A_0$, which is a scalar on $\S^3$, 
the Ricci tensor part is not present and $\nabla^2$ is equivalent to
the scalar Laplacian. The eigenvectors of the scalar Laplacian are
spherical harmonics $Y_\ell$ 
labelled by angular momentum quantum numbers, $\ell\in{\mathbb Z}\geq0$ with 
\EQ{
\Delta Y_\ell=-R^{-2}\ell(\ell+2)Y_\ell
}
and degeneracy
$(\ell+1)^2$. The vector field $A_i$ can be decomposed 
into the image and the kernel of
: $A^i=B^i+C^i$ with $\nabla_iB^i=0$ and 
$C^i=\nabla^i f$. These modes have different eigenvalues for the
vector Laplacian.
Firstly, those in the image of $\nabla^i$, {\it
  i.e.\/}~$C^i$, are given by $\nabla^iY_\ell$ with $\ell\in{\mathbb Z}>0$, 
which satisfy
\EQ{
\Delta\nabla^iY_\ell=-R^{-2}\ell(\ell+2)\nabla^iY_\ell\ .
}
The remaining modes $B^i$, in the kernel of $\nabla^i$, 
are spanned by $V_\ell^i$, also labelled by the
angular momentum $\ell\in{\mathbb Z}>0$, with
\EQ{
\Delta V^i_\ell=-R^{-2} (\ell+1)^2V^i_\ell
}
and degeneracy $2\ell(\ell+2)$. Finally, we will need the Laplacian on
fermionic modes. For 2-component real fermions\footnote{Note that a
  Weyl fermion in four dimensions corresponds to two 2-component real
  fermions.} on $\S^3$, the
fermionic Laplacian has eigenvalues $R^{-2}(\ell+\tfrac12)$, with
$\ell\in{\mathbb Z}>0$ and degeneracy $\ell(\ell+1)$.

The gauge field modes $B_i$ and $C_i$ decouple 
from the chemical
potential and at Gaussian order the relevant terms in the Lagrangian
are 
\EQ{
\frac12\text{Tr}\,B_i\big[-\tilde D_0^2-\Delta^{(v)}+\varphi^2\big]B^i+
\frac12\text{Tr}\,B_i\big[-\tilde D_0^2-\Delta^{(s)'}+\varphi^2\big]B^i
\ ,
}
where the superscript reminds us that the Laplacian is for 
divergenceless vectors and scalars, respectively. The prime means that
the $\ell=0$ mode is missing. 
When these fluctuations are integrated out their
contribution to the effective potential is of the form 
\EQ{
\frac1{2\beta}\sum_{ij=1}^N\sum_{\ell=\ell_0}^\infty d_\ell \log\det\big[-
\tilde D_0^2+\varepsilon_\ell(\varphi_{ij})^2\big]\ ,
}
where $d_\ell$ is the degeneracy of the modes with angular momentum
quantum number
$\ell$, so $2\ell(\ell+2)$ for $B_i$ and $(\ell+1)^2$ for $C_i$. The
integer $\ell_0$ is the lower limit
on the angular momentum quantum number, so $1$ for both $B_i$ and
$C_i$. The quantity
$\varepsilon_\ell$ is the energy of the mode, so
equal to 
\SP{
B_i:&\qquad\varepsilon_\ell=\sqrt{R^{-2}(\ell+1)^2+\varphi^2}\
,\\
C_i:&\qquad\varepsilon_\ell=\sqrt{R^{-2}\ell(\ell+2)^2+\varphi^2}\
\ .
}
The contributions from all the other modes have the same form for some
set of $\{d_\ell,\ell_0,\varepsilon_\ell\}$.

\end{document}